\documentclass[aip,jcp,reprint,amsmath,amssymb]{revtex4-2}
\usepackage{graphicx}
\usepackage[suffix=]{epstopdf}
\usepackage{natmove}
\usepackage{amsmath,amssymb}

\newcommand{\vv}[1]{\mathbf{#1}}
\renewcommand{\d}[1]{{\operatorname{d}\!{#1}}}

\begin{document}
\title{Mesoscale particle-based simulations of flow in expansion--contraction microchannels at low Reynolds number}

\author{Tzortzis Koulaxizis}
\thanks{These authors contributed equally.}
\affiliation{Department of Chemical and Biomolecular Engineering, University of Illinois, Urbana--Champaign, Illinois 61801, USA}

\author{Clara De La Torre Garcia}
\thanks{These authors contributed equally.}
\affiliation{Department of Chemical Engineering, Auburn University, Auburn, AL 36849, USA}

\author{C. Levi Petix}
\affiliation{Department of Chemical Engineering, Auburn University, Auburn, AL 36849, USA}

\author{Antonia Statt}
\email{statt@illinois.edu}
\affiliation{Department of Chemical and Biomolecular Engineering, University of Illinois, Urbana--Champaign, Illinois 61801, USA}
\affiliation{Department of Materials Science and Engineering, The Grainger College of Engineering,  University of Illinois, Urbana--Champaign, Illinois 61801, USA}

\author{Michael P. Howard}
\email{mphoward@auburn.edu}
\affiliation{Department of Chemical Engineering, Auburn University, Auburn, AL 36849, USA}

\begin{abstract}
We computationally study the flow of Newtonian fluids through sinusoidal expansion--contraction microchannels at low Reynolds number. We first use a perturbation method to analytically derive series solutions for the stream function and volumetric flow rate that extend prior work [P.K.~Kitanidis and B.B.~Dykaar, \textit{Transport in Porous Media} \textbf{26}, 89--98 (1997)] up to tenth order. We then employ two particle-based mesoscale methods, dissipative particle dynamics (DPD) and multiparticle collision dynamics (MPCD), to simulate the same flows. We find that the fluid velocity at the expansion and contraction points as well as the volumetric flow rate are in good agreement between DPD, MPCD, and the fourth-order series solution for a wide range of microchannel geometries. The mesoscale fluid models exhibit some slip at the walls, leading to a small but consistent overprediction of the velocity and volumetric flow rate. The series solution fails for short microchannel lengths and large amplitudes; we identify lengths and amplitudes for which it converges to a given order. Overall, we find that DPD and MPCD are convenient and reasonably accurate methods, particularly for microchannel geometries where the series solution fails or is cumbersome to implement.  
\end{abstract}

\maketitle

\section{Introduction}
Flow through expansion--contraction microchannels is used in a variety of technologies ranging from microfluidic mixers\cite{Mondal2019} to particle separators \cite{review_contrexp, lee_inertial_sep,blood_sep}. For example, expansion--contraction microchannels have been used to study DNA translocation \cite{dna_in_micro,mai_schroeder_review, dna_microchannel_conv} and the filtration of polymers with different architectures \cite{Locatelli_2023}. The curvature and constriction of expansion--contraction microchannels can trigger complex transport phenomena; for example, pulsatile forcing of a Newtonian fluid in a sinusoidal channel enhances mass transfer \cite{nishimura_pulsatile}. Expansion--contraction microchannels also serve as platforms for studying transport in porous media\cite{viscoelastic_porous_padding,jagdale_fluid_contr_exp,Browne2020b}, especially the rich behaviors of non-Newtonian fluids in these geometries \cite{datta_perspectives_2022} that are relevant for processes such as enhanced oil recovery \cite{Xie2022,single_phase_clarke,viscoelastic_clarke}.

Both numerical and analytical methods have been applied to study flow in expansion--contraction geometries. Boundary integral methods were used to identify flow-reversal criteria and quantify geometric effects for Newtonian fluids in periodic constriction channels \cite{Pozrikidis1987}. Pressure drop corrections and criteria for recirculation based on the amplitude-to-length ratio were derived with a similar method \cite{hemmat_creeping}. Lattice Boltzman simulations also have been applied to simulate Newtonian fluids in expansion--contraction geometries \cite{velasquez_ortega_bingham,ortega_friction_newtonian}. Analytical perturbation methods were used to model Newtonian fluids in two-dimensional periodic pores \cite{Kitanidis1997} and three-dimensional wavy walls \cite{malevich}. Non-Newtonian fluids are more challenging to describe with analytical approaches, but Boyko and Stone recently analyzed the behavior of Oldroyd-B type models in contractions using lubrication theory \cite{Boyko_Stone_2022,Boyko_Hinch_Stone_2024, Hinch_Boyko_Stone_2024_contraction}. They also provide an excellent summary of prior numerical works on flow of non-Newtonian fluids in non-uniform geometries \cite{Boyko_Stone_2022}.

Mesoscale particle-based simulation methods such as dissipative particle dynamics (DPD) \cite{groot1997} and multiparticle collision dynamics (MPCD) \cite{howard_modeling_2019} are intriguing alternatives for simulating microchannel flows. DPD represents a fluid as soft particles that interact through pairwise conservative, dissipative, and random forces, while MPCD represents a fluid as point-like particles that periodically participate in stochastic momentum-exchanging collisions. Both can efficiently model not only Newtonian fluids but also non-Newtonian fluids, e.g., by explicitly embedding coarse-grained polymers into a solvent. Such mesoscale representations may be critical for accurately describing transport phenomena for non-Newtonian fluids \cite{KOPPOL_SURESHKUMAR_ABEDIJABERI_KHOMAMI_2009}.
DPD has been used to study the trapping of fluid particles \cite{kasiteropoulou_grooved_nanochannels} and transport of macromolecules\cite{zhou2013grooved} in grooved nanochannels,
flow of DNA in periodic contraction--diffusion channels\cite{fan2006}, and polymer translocation dynamics \cite{dpd_translocation_polymer,translocation_nanochannel_star_zhu}; while MPCD has been used to investigate flow of colloidal particles in porous environments \cite{colloid_microfluidic}, translocation of knotted polymers \cite{translocation_mpcd_knot}, and filtration of ring polymers in expansion--contraction microchannels \cite{Weiss2019}.

However, despite their popularity, to our knowledge there is no systematic study on whether DPD and MPCD accurately describe even the flow of Newtonian fluids in expansion--contraction microchannels. Establishing accuracy for this baseline case is important for using mesoscale methods to reliably simulate more complex phenomena, such as those for non-Newtonian fluids. In this work, we interrogate the ability of DPD and MPCD to model the flow of Newtonian fluids through sinusoidal expansion--contraction microchannels at low Reynolds number. The mesoscale particle-based simulations are compared to an approximate analytical theory. Overall, we find good agreement between both particle-based simulation methods and the theory for microchannel geometries for which the theory is expected to be accurate. Importantly, though, the particle-based methods can also be used to simulate flows in microchannels for which the theory fails.

The rest of this article is organized as follows. Section \ref{sec:theory} presents the sinusoidal expansion--contraction microchannel geometry that we studied, along with a theoretical solution based on a perturbation method, while Section \ref{sec:simulation} describes the particle-based simulation models and methods. The simulation results are critically compared to theoretical predictions in Section \ref{sec:results}; we examine the fluid velocity and volumetric flow rate in different microchannels, and we discuss shortcomings of both modeling approaches. Section \ref{sec:Conclusions} summarizes the main findings and provides an outlook.

\section{Theory}
\label{sec:theory}
\begin{figure}
    \centering
    \includegraphics{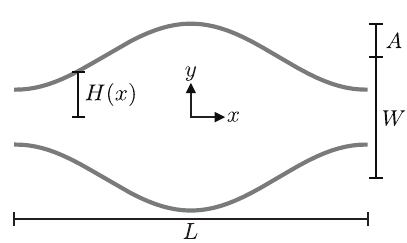}
    \caption{Schematic of expansion--contraction microchannel geometry.}
    \label{fig:channel}
\end{figure}

We considered an expansion--contraction microchannel that is effectively two-dimensional with a net flow in the $x$ direction and boundary walls in the $y$ direction (Figure \ref{fig:channel}). The half width between the walls $H$ varied sinusoidally as 
\begin{equation}
H(x) = \frac{W}{2}+A\cos\left(\frac{2\pi x}{L}\right),
\label{eq:walls}
\end{equation}
where $L$ is the microchannel length, $W$ is the average distance between the walls, and $A$ is the amplitude of the walls. The distance between the microchannel walls at the expansion and contraction points was hence $W \pm 2A$. The microchannel nominally also had depth $D$ in the $z$ direction, and there were periodic boundary conditions in the $x$ and $z$ directions. The domain for the fluid was taken to be $-L/2 \le x < L/2$, $-H(x) \le y \le H(x)$, and $-D/2 \le z < D/2$.

Kitanidis and Dykaar \cite{Kitanidis1997} studied the flow of a Newtonian fluid at low Reynolds numbers through a similar geometry and obtained a fourth-order series solution in the parameter $\epsilon = W/L$. Their approach is based on a stream function formulation of the Stokes equations, which they solved using a perturbation method \cite{deen2012analysis}. The solution is expected to be accurate when the channel is long compared to the average wall separation, but they did not test its range of validity. Their derivation also has some notational inconsistencies and missing details. Here, we clarify how to solve for the flow in this microchannel geometry using their approach, and we extend their solution to tenth order for reasons that will be discussed in Section \ref{sec:results}.

We assumed a steady incompressible flow of a Newtonian fluid with viscosity $\mu$ at low Reynolds number, so the fluid velocity $\vv{u}$ was governed by the Stokes equations \cite{Guazzelli_Morris_Pic_2011},
\begin{align}
\nabla \cdot \vv{u} &= 0 \label{eq:continuity} \\
\nabla \cdot \boldsymbol{\sigma} &= \vv{0}, \label{eq:stokes}
\end{align}
where $\boldsymbol{\sigma} = -p\boldsymbol{\delta} + \boldsymbol{\tau}$ is the stress tensor, $p$ is the pressure, $\boldsymbol{\delta}$ is the identity tensor, and $\boldsymbol{\tau} = \mu[\nabla\vv{u} + (\nabla\vv{u})^{\rm T}]$ is the deviatoric stress tensor. Written this way, the pressure may also include a conservative body force on the fluid. The boundary conditions were no slip and no penetration at the walls so $\vv{u} = \vv{0}$ at $y = \pm H$, periodicity in $x$ so $\vv{u}$ is equal at $x = \pm L/2$, and a constant pressure drop $\Delta p > 0$ between $x = -L/2$ and $x = L/2$ to drive the flow in the positive $x$ direction. By symmetry, there should be no flow in the $z$ direction ($u_z = 0$ everywhere), and $\vv{u}$ should not depend on $z$. 

Kitadanis and Dyakaar \cite{Kitanidis1997} reformulated the pressure boundary condition as an auxiliary condition on the volumetric flow rate $Q$ by balancing the rate of work done on the fluid with the rate of viscous dissipation $\mu \Phi = \boldsymbol{\tau} : \nabla\vv{u}$ \cite{Winter1987},
\begin{equation}
\int_S \vv{n} \cdot \boldsymbol{\sigma}\cdot\vv{u} \d{\vv{x}} = \int_V \mu \Phi \d{\vv{x}},
\label{eq:workdissip}
\end{equation}
where $S$ is the closed boundary of the microchannel volume $V$ with outer normal $\vv{n}$. We expected $\boldsymbol{\tau}$ to be periodic because it is a function of $\vv{u}$. Along with the boundary conditions and the integral definition of the volumetric flow
\begin{equation}
Q = \int_{A_{\rm c}} \vv{n} \cdot \vv{u} \d{\vv{x}}
\label{eq:volumetric}
\end{equation}
through a cross-sectional area $A_{\rm c}$, the first integral was evaluated to be $\Delta p Q$. The second integral depends on the velocity $\vv{u}$ through $\Phi$.

The Stokes equations can be reformulated using a stream function $\psi(x,y)$ such that $\vv{u} = \nabla\times(\psi\vv{e}_z)$, where $\vv{e}_z$ is the basis function in the $z$ direction. This stream function must satisfy the biharmonic equation $\nabla^4\psi = 0$, and the boundary conditions on the velocity become boundary conditions on the corresponding first derivatives of $\psi$. For our microchannel geometry, $Q$ is also related to the stream function by Stokes' theorem as $Q = [\psi(x,H)-\psi(x,-H)] D$. The stream function is only defined up to a constant so we arbitrarily chose $\psi(x, -H) = -Q/(2D)$, which implies $\psi(x, H) = Q/(2D)$.

To solve the biharmonic equation for $\psi$, we nondimensionalized the equations as follows: lengths were scaled by $W$, the stream function was scaled by $W U_\parallel$ where
\begin{equation}
U_\parallel = \frac{W^2}{12 \mu} \frac{\Delta p}{L}
\label{eq:uparallel}
\end{equation}
is the average velocity in a parallel-plate microchannel, and the volumetric flow rate was scaled by that of the parallel-plate microchannel, $Q_\parallel = W D U_\parallel$. All quantities in this section made dimensionless in this way will be denoted using a hat. Explicitly, the biharmonic equation became
\begin{equation}
\frac{\partial^4 \hat\psi}{\partial \hat x^4} + 2 \frac{\partial^4 \hat\psi}{\partial \hat x^2 \partial \hat y^2} + \frac{\partial^4 \hat\psi}{\partial \hat y^4} = 0,
\label{eq:biharmonic}
\end{equation}
and we also had
\begin{equation}
\hat \Phi = 4 \left(\frac{\partial ^2 \hat \psi}{\partial \hat x \partial \hat y}\right)^2 + \left(\frac{\partial ^2 \hat \psi}{\partial \hat y^2}-\frac{\partial^2 \hat \psi}{\partial \hat x^2} \right)^2.
\label{eq:phi}
\end{equation}
Next, we transformed the boundary conditions to use rescaled coordinates $\eta = \epsilon \hat x$ and $\xi = \hat y / \hat H$ having a rectangular domain $-1/2 \le \eta < 1/2$ and $-1 \le \xi \le 1$. In terms of these variables, the half width between the walls became
\begin{equation}
\hat H(\eta) = \frac{1}{2} + \hat A \cos(2\pi \eta),
\end{equation}
and Eq.~\eqref{eq:workdissip} became
\begin{equation}
\hat{Q} = \frac{1}{12}\int_{-1/2}^{1/2} \int_{-1}^1 \hat{\Phi}\hat{H}d\xi d\eta \label{eq:aux}.
\end{equation}
Some care must be taken in transforming the derivatives of $\hat \psi$ to use these coordinates. For example,
\begin{subequations}
\begin{align}
\frac{\partial\hat\psi}{\partial \hat x} &= \epsilon \left(\frac{\partial \hat\psi}{\partial \eta} - \xi \frac{\hat H_1}{\hat H} \frac{\partial \hat \psi}{\partial \xi}\right) \\
\frac{\partial\hat\psi}{\partial \hat y} &= \frac{1}{\hat H} \frac{\partial \hat \psi}{\partial \xi},
\end{align}
\end{subequations}
where $\hat H_i$ designates the $i$-th derivative of $\hat H$ with respect $\eta$. Similar derivative transformations must be applied to all terms in Eqs.~\eqref{eq:biharmonic} and \eqref{eq:phi}. The boundary conditions on $\hat \psi$ were $\hat \psi = \pm \hat Q/2$, $\partial\hat\psi/\partial\xi = 0$, and $\partial\hat\psi/\partial\eta = 0$ at $\xi = \pm 1$; and $\partial\hat\psi/\partial\xi$ and $\partial\hat\psi/\partial\eta$ being periodic at $\eta = \pm 1/2$.

An approximate solution to the biharmonic equation with these boundary conditions can be derived using a perturbation method with the small parameter $\epsilon$, i.e., microchannels that are long compared to their average width \cite{Kitanidis1997}. We sought solutions of the form
\begin{equation}
\hat \psi(\eta,\xi) = \sum_{n=0}^\infty \hat\psi_n(\eta,\xi) \epsilon^n,
\end{equation}
where $\{\hat \psi_n\}$ are unknown functions to be determined. Given the form of Eq.~\eqref{eq:aux}, we also expected
\begin{equation}
\hat Q = \sum_{n=0}^\infty \hat Q_n \epsilon^n
\end{equation}
to have a corresponding series, where $\{\hat Q_n\}$ are unknown coefficients to be determined. The series for $\hat \psi$ was substituted in the biharmonic equation, then like-powers of $\epsilon$ were collected to obtain a set of partial differential equations for $\{\hat \psi_n\}$. These partial differential equations could be solved using only the boundary conditions $\partial\hat\psi_n/\partial\xi = 0$ at $\xi = \pm 1$ and $\hat\psi_n = \pm \hat Q_n/2$ at $\xi = \pm 1$. The solutions for $\{\hat \psi_n\}$ contained the unknown coefficients $\{\hat Q_n\}$, which we subsequently determined by substituting $\{\hat \psi_n\}$ in Eq.~\eqref{eq:aux} and solving a system of linear equations. We performed these calculations using Mathematica version 14.2.1.0, which enabled us to obtain a solution up through tenth order.

The solution contained only even powers of $\epsilon$. The terms through fourth order were
\begin{subequations}
\begin{align}
        \hat\psi_{0} &= \hat{Q}_{0} \hat f_{0} \\
        \hat\psi_{2} &= \hat{Q}_{2} \hat f_{0} + \hat{Q}_{0} \hat f_{2}  \\
        \hat\psi_{4} &= \hat{Q}_{4} \hat f_{0} + \hat{Q}_{2} \hat f_{2} + \hat{Q}_{0} \hat f_{4}
\end{align}
\end{subequations}
with functions $\{\hat f_n(\eta,\xi)\}$,
\begin{subequations}
\begin{align}
\hat f_{0} &= -\frac{1}{4} \xi \big( \xi^2 - 3 \big) \\
\hat f_{2} &= \frac{3}{40} \xi \big( \xi^2 - 1 \big)^2 \big( 4 \hat{H}_1^2 - \hat{H} \hat{H}_2 \big) \\
\hat f_{4} &= -\frac{1}{5600}
    \xi \big( \xi^2 - 1 \big)^2 \Big[
        24 \left( 75 \xi^2 + 17 \right) \hat{H}_1^4 \nonumber \\
&+ \hat{H}^2 \Big(
            \left( 19 - 15 \xi^2 \right) \hat{H} \hat{H}_4
            + 90 \left( 2 \xi^2 - 3 \right) \hat{H}_2^2
        \Big) \nonumber \\
&+ 8 \left( 30 \xi^2 - 31 \right) \hat{H}^2 \hat{H}_3 \hat{H}_1 \nonumber \\
&- 36 \left( 50 \xi^2 - 19 \right) \hat{H} \hat{H}_1^2 \hat{H}_2
    \Big],
\end{align}
\end{subequations}
and coefficients
\begin{subequations}
\begin{align}
\hat{Q}_{0} &=\frac{1}{\hat{I}_{0}} \\
\hat{Q}_{2} &=\frac{\hat{I}_{2}}{\hat{I}_{0}^2} \\
\hat{Q}_{4} &=\frac{\hat{I}_{2}^2-\hat{I}_{0} \hat{I}_{4}}{\hat{I}_{0}^3},
\end{align}
\end{subequations}
along with the integrals
\begin{subequations}
\begin{align}
\hat{I}_{0} &=\frac{1}{8} \int_{-1/2}^{1/2} \hat{H}^{-3} \d{\eta} \\
\hat{I}_{2} &= \frac{1}{20} \int_{-1/2}^{1/2} 
        \big( \hat{H} \hat{H}_2 - 5 \hat{H}_1^2 \big) 
        \hat{H}^{-3} \d{\eta} \\ 
\hat{I}_{4} &= \frac{1}{1400}  \int_{-1/2}^{1/2}
        \big(
            -4 \hat{H}^3 \hat{H}_{4}
            + 54 \hat{H}^2 \hat{H}_2^2
            + 87 \hat{H}_1^4 \nonumber \\
            &\quad+ 56 \hat{H}^2 \hat{H}_{3} \hat{H}_1
            - 306 \hat{H} \hat{H}_1^2 \hat{H}_2
        \big) 
        \hat{H}^{-3} \d{\eta}.
\end{align}
\end{subequations}
We verified these results were equivalent to those of Kitanidis and Dyakaar with different notation\cite{Kitanidis1997}. We present complete details of the calculations through tenth order in the Supplementary Material.

\section{Simulation}
\label{sec:simulation}
We simulated the same microchannel geometry described in Section \ref{sec:theory} using two mesoscale particle-based methods: DPD \cite{original_dpd, espanol1995} and MPCD \cite{malevanets_1999_mesoscopic}. To facilitate comparison of these methods, all simulation parameters and results will be presented using a consistent system of units having unit of mass $m$, unit of length $\ell$, and unit of energy $\varepsilon$. The unit of time is $\tau = \sqrt{m\ell^2/\varepsilon}$. All particles had mass $M = 1.0\,m$, and all simulations were conducted at constant temperature $T = 1.0\,\varepsilon/k_{\rm B}$, where $k_{\rm B}$ is the Boltzmann constant. Note that $\varepsilon$ used here and later in this article as the unit of energy should not be confused with $\epsilon = W/L$ in Section \ref{sec:theory}.

Simulations were performed for microchannel lengths $L$ between $30\,\ell$ and $150 \ell$. The mean width between walls $W$ and the depth $D$ were fixed at $30\,\ell$, and the wall amplitude $A$ was varied between $0\,\ell$ and $12\,\ell$. Periodic boundary conditions were applied in the $x$ and $z$ directions, and a bounce-back rule was used to model no-slip and no-penetration boundary conditions at the walls in the $y$ direction \cite{Lamura2001b, Whitmer_2010}. Intersection of particles with the walls was computed numerically using a combination of Newton's method and bisection search, and the velocities of particles that intersected with the wall were reflected. The orthorhombic simulation box containing the microchannel was padded by $4\,\ell$ in both the positive and negative $y$ direction at the expansion point to accommodate wall particles for DPD \cite{revenga1999,pivkin2005} and virtual particles \cite{Lamura2001b,Whitmer_2010} for MPCD (see below). All simulations were performed using HOOMD-blue \cite{hoomd, PHILLIPS,howard_efficient_2018} (version 5.3.0) extended with azplugins \cite{azplugins} (version 1.1.0). We present specific details of the two simulation methods next.

\subsection{Dissipative particle dynamics}
DPD uses particles that have three types of pairwise interactions: a conservative force, a dissipative force, and a random force \cite{original_dpd, espanol1995, groot1997}. In our simulations, all particles interacted through the standard DPD conservative repulsive force with interaction parameter $25\,\varepsilon/\ell$, but we used a modified weight function with exponent $1/2$ and friction coefficient $4.5\,m/\tau$ for the dissipative and random forces to produce a more liquid-like fluid \cite{fan2006,howard2019}. The pairwise distance cutoff for all forces was $1.0\,\ell$, and the equations of motion for the particles were integrated using a velocity Verlet scheme\cite{Allen} with timestep $0.01\,\tau$.

For each microchannel geometry, particles were first randomly dispersed everywhere in the simulation box with no walls and three-dimensional periodic boundary conditions at a number density of $\rho_{\rm n} = 3/\ell^3$. This bulk fluid was equilibrated for $10^{3}\tau$ then converted to represent a fluid confined between walls. Particles inside the microchannel continued to represent the fluid [blue particles in Fig.~\ref{fig:channel}(a)], and their center-of-mass velocity was set to zero. Particles outside the microchannel but within a distance of $3\,\ell$ in the $y$ direction were converted to be wall particles [gray particles in Fig.~\ref{fig:channel}(a)], their velocities were set to zero, and their motion was no longer integrated. All other particles were deleted. The presence of explicit wall particles helped reduce slip and undesirable density oscillations compared to using only bounce-back reflection \cite{revenga1999,fedosov2008,pivkin2005,howard2019}. The fluid particles were then equilibrated for an additional $10^3\,\tau$ before starting a flow simulation.

\begin{figure}
    \centering
    \includegraphics{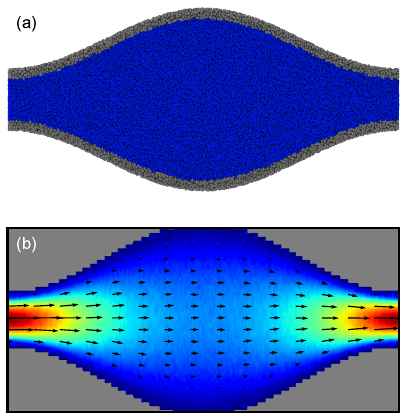}
    \caption{DPD simulation of flow in expansion--contraction microchannel with $L=150\,\ell$, $W=30\,\ell$, and $A=9 \,\ell$. (a) Snapshot of the microchannel with fluid particles (blue) and wall particles (gray). Image rendered using VMD (version 1.9.4) \cite{humphrey:jmolgraph:1996}. (b) Relative magnitude and direction of the simulated fluid velocity in the same microchannel, with red being fastest and blue being slowest. Figure S1 shows the equivalent of (b) for the MPCD fluid.}
    \label{fig:simulation}
\end{figure}

\subsection{Multiparticle collision dynamics}
MPCD uses particles that do not interact through pairwise forces; instead, their positions and velocities evolve through alternating streaming and collision steps \cite{malevanets_1999_mesoscopic,gompper_2008_mpcd_review,howard_modeling_2019}. In our simulations, the particles moved ballistically for $0.1\,\tau$ during the streaming step. Their positions and velocities were updated using a velocity Verlet scheme \cite{Allen}, and bounce-back reflection was enforced with the walls \cite{Lamura2001b,Whitmer_2010}. During the collision step, the particles were binned into cubic cells of edge length $1.0\,\ell$ then exchanged momentum with other particles in the same cell using the stochastic rotation dynamics scheme \cite{malevanets_1999_mesoscopic} with a fixed rotation angle of $130^\circ$, a rotation axis randomly drawn from the unit sphere for each cell, and a cell-level Maxwell--Boltzmann thermostat \cite{huang_thermostat_mpcd}. Following best practices, the cells were shifted in each Cartesian direction by a uniform random amount between $-\ell/2$ and $\ell/2$ before each collision \cite{ihle_2001_galilean_srd}. Virtual particles were also added everywhere outside the walls using random sampling with rejection and participated in the collision \cite{Lamura2001b,Whitmer_2010}.

For each microchannel geometry, MPCD particles were randomly inserted between the walls at number density $\rho_{\rm n} = 10/\ell^3$, and a flow simulation was immediately commenced.

\subsection{Simulation protocol}
To simulate flow with both methods, a constant force $F$ was applied to all fluid particles in the $x$ direction \cite{Allahyarov2002}. In selecting the value of $F$, we desired to have a small Reynolds number ${\rm Re} = \rho U_\parallel W/\mu$, where $\rho = M\rho_{\rm n}$ is the mass density of the fluid and $U_\parallel$ is calculated using Eq.~\eqref{eq:uparallel} with $\Delta p/L = \rho_{\rm n} F$, for consistency with physical microchannel flows and the assumption underlying our theoretical analysis (Section~\ref{sec:theory}) yet a large enough velocity to be reliably measured. We used $F = 10^{-3}\,\varepsilon/\ell$ in the DPD simulations and $F = 2.5 \times 10^{-3}\,\varepsilon/\ell$ in the MPCD simulations, which gave ${\rm Re} \approx 7.6$ and 7.5, respectively, based on the viscosities we calculated (Section \ref{sec:results}). In each simulation, the flow was allowed to develop for $10^4\tau$, then the fluid velocity $\vv{u}$ and volumetric flow rate $Q$ were measured using the procedures described below during a production period of $10^5\tau$.

The fluid velocity at the expansion and contraction points $u_x(y)$ was measured using histograms sampled every $10\,\tau$. Fluid particles with an $x$ coordinate within a thickness $0.5\,\ell$ centered at the expansion were binned with respect to their $y$ coordinate between the walls using a bin size of $0.5\,\ell$, and the velocities of particles in each bin were averaged. The same procedure was used at the contraction, but due to the periodic boundary conditions, particles were binned in two separate half-thicknesses at each end of the simulation box and the resulting histograms were averaged together with equal weight. The two-dimensional velocity field $\vv{u}(x,y)$ was also calculated by binning the entire simulation box with respect to the $x$ and $y$ coordinates of the fluid particles using the same bin size and sampling frequency.

The volumetric flow rate $Q(x)$ at a given position $x$ can be calculated using Eq.~\eqref{eq:volumetric} by integrating $\vv{u}(x,y)$ over the cross-sectional area $A_{\rm c}(x) = 2 H(x) D$. However, the MPCD fluid can exhibit significant density variations in response to pressure variations (Figure S2) because it has an ideal-gas equation of state \cite{malevanets_1999_mesoscopic, srd_gk,Howard_hydrodynamics_colloid,stark_eos}. Variations in density should lead to variations in $Q$ because of mass conservation, so a volumetric flow rate averaged over the microchannel may be more appropriate than averaging only at a particular point (e.g., the expansion or contraction). Accordingly, we instead computed the volumetric flow rate using an ensemble average over the particles.

Based on Eq.~\eqref{eq:volumetric}, we defined the average volumetric flow rate as
\begin{equation}
Q = \int Q(x) f^{(1)}(x)\d{x} = \int A_{\rm c}(x) U(x) f^{(1)}(x)\d{x},
\end{equation}
where $f^{(1)}(x)$ is the one-particle distribution function for finding a particle at position $x$\cite{theory_simple_liq}, and $U(x)$ is the average $x$-component of the fluid velocity through the cross-sectional area $A_{\rm c}(x)$ at $x$. We note that there is not a unique definition of this average and others are possible; we used this one because it more heavily weights $x$ positions where fluid is more likely to be found. We then used the one-particle  distribution function $f^{(1)}(x, y, z, v_x)$ for finding a particle at position $(x,y,z)$ with $x$-velocity $v_x$ to express the average velocity as
\begin{equation}
U(x)= \frac{\int v_x f^{(1)}(x,y,z,v_{x}) \d{y} \d{z} \d{v_x}}{\int f^{(1)}(x,y,z,v_{x}) \d{y} \d{z} \d{v_x}}.
\end{equation}
Using the fact that
\begin{equation}
f^{(1)}(x) = \int f^{(1)}(x,y,z,v_x) \d{y}\d{z}\d{v_x}
\end{equation}
and simplifying gave the ensemble average
\begin{equation}
Q = \langle A_{\rm c}(x) v_x\rangle.
\end{equation}
We performed this average in our simulation by sampling the particle coordinates every $10^2\,\tau$.

\section{Results and Discussion}
\label{sec:results}
We compared the fluid velocity and volumetric flow rates simulated using DPD and MPCD (Section~\ref{sec:simulation}) with those calculated theoretically (Section~\ref{sec:theory}). To make this comparison, we required the viscosity $\mu$ of the DPD and MPCD fluids. Expressions for $\mu$ in terms of model parameters exist for both methods \cite{ripoll_kin_coll_visc, fan2006}; the expression for MPCD is known to be quite accurate \cite{statt_unexpected_2019}, but the expression for DPD is less accurate \cite{howard2019}. Accordingly, we determined $\mu$ for both methods using our simulations.

To use as much of our data as possible, we averaged the two-dimensional velocity in two parallel-plate microchannels ($A = 0\,\ell$, $L =  75\,\ell$ and $150\,\ell$) with respect to $x$ using equal weight to obtain $u_x(y)$ (Figure S3). We expected a parabolic flow $u_x(y) = (3U_\parallel/2)[1-(2y/W)^2]$ in this geometry for no-slip boundary conditions. However, to avoid ambiguities in determining the viscosity if there is wall slip, we used the velocity gradient $\d{u_x}/\d{y} = -12U_\parallel y/W^2$. We numerically differentiated the measured $u_x$ with respect to $y$; it was mostly linear, as expected, but there were small deviations near the walls (Figure S4). We fit $\d{u_x}/\d{y}$ for $|y| \le 10\,\ell$ to a linear form, and we extracted viscosities of $1.63\,\varepsilon\tau/\ell^3$ for DPD and $8.68\,\varepsilon\tau/\ell^3$ for MPCD. The viscosity for DPD is in good agreement with previous measurements using reverse-nonequilibrium simulations \cite{howard2019} but is about 14\% larger than the theoretical prediction of $1.43\,\varepsilon\tau/\ell^3$ \cite{fan2006}, while the viscosity for MPCD is in excellent agreement with the theoretical expectation of $8.70\,\varepsilon \tau/\ell^3$ \cite{ripoll_kin_coll_visc}. We used the values we measured to normalize the simulation results in all subsequent comparisons.

\subsection{Velocity}
\begin{figure}
    \centering
    \includegraphics{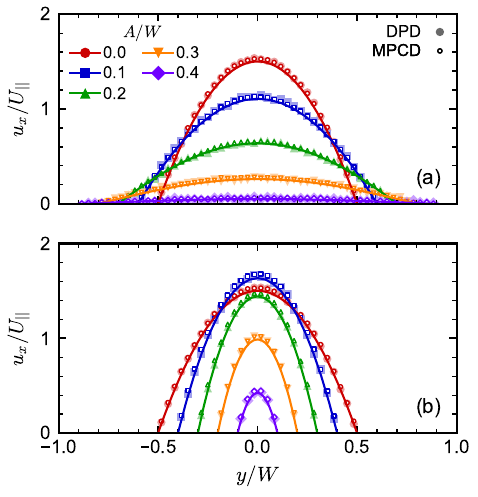}
    \caption{Velocity $u_x(y)$ at the (a) expansion and (b) contraction for various amplitudes $A$ at fixed length $L/W = 5$. The velocity is normalized by the theoretically expected average velocity between parallel plates, $U_\parallel$. Closed symbols are DPD simulations, open symbols are MPCD simulations, and lines are theoretical predictions using the fourth-order solution.}
    \label{fig:amp_dependence_1}
\end{figure}

We first compared the simulated fluid velocities at the expansion and contraction points to the fourth-order theoretical predictions for microchannels with fixed length $L/W = 5$ and amplitudes varying from $A/W = 0$ to 0.4 (Figure \ref{fig:amp_dependence_1}). The theoretical predictions for the velocity were obtained by analytically differentiating the stream function. The maximum velocity at the expansion systematically decreased as $A$ increased; at the contraction, it initially increased before decreasing. In all cases, the DPD and MPCD simulations were in nearly quantitative agreement, demonstrating consistency of the methods, but both gave velocities that were slightly larger than the theoretical predictions (Figure \ref{fig:amp_dependence_1} and S5). We expected the theoretical predictions to be accurate for these microchannel geometries because they were long compared to their width ($W/L = 0.2$ is reasonably small). We also saw a comparable amount of disagreement between the simulations and theory even for the parallel-plate microchannel, for which the theoretical prediction is exact. We hence attribute the difference between the simulations and theory to the presence of a small amount of effective wall slip.

In support of this notion, we refit the velocities in the parallel-plate microchannels that we used to determine the viscosity, but we now allowed for a finite slip velocity at the walls (Figure S3). The fit slip velocity was 2.3\% of $U_\parallel$ for DPD and 3.2\% for MPCD. This amount of slip is consistent with deviations in the velocity for all microchannels in Figure \ref{fig:amp_dependence_1}. Some possible origins of the effective slip might be imperfect enforcement of no-slip boundary conditions using the bounce-back reflection scheme, as well as small deviations in fluid properties near the walls. However, the agreement between simulations and theory is overall very good, validating the implementation of both simulation models for the expansion--contraction microchannel as well as the theoretical calculations.

\subsection{Volumetric flow rate}
\begin{figure}
    \centering
    \includegraphics{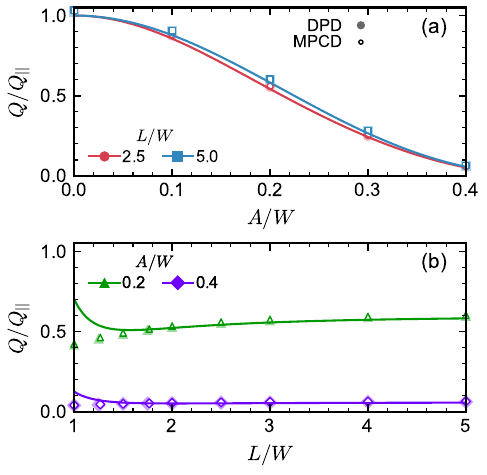}
    \caption{Volumetric flow rate $Q$ as a function of (a) amplitude $A$ with fixed length $L$ and (b) length $L$ with fixed amplitude $A$. The volumetric flow rate is normalized by the expected volumetric flow rate between parallel plates, $Q_\parallel$. Closed symbols are DPD simulations, open symbols are MPCD simulations, and lines are fourth-order theoretical predictions.}
    \label{fig:Q_A}
\end{figure}

Given the good agreement in the velocity between the simulations and theory for long microchannels ($L/W = 5$), we next tested whether similarly good agreement could be obtained in shorter microchannels. We halved the length to $L/W = 2.5$, and we conducted simulations for the same amplitudes as in Figure \ref{fig:amp_dependence_1}. In order to more easily visualize the results for multiple microchannels, we will now focus on the volumetric flow rate $Q$ rather than the velocity [Figure \ref{fig:Q_A}(a)]. For both lengths, $Q$ decreased as the amplitude increased due to the enhanced constriction of the flow, reaching $Q/Q_\parallel \approx 0.1$ for the most constricted channels. Both the DPD and MPCD simulations were again in good agreement with the fourth-order theoretical predictions, with the simulated values being between 2.5\% (for $A/W = 0$) and 8.5\% (for $A/W = 0.4$) larger than the theoretical predictions when $L/W = 5$. This overprediction is consistent with the effective slip we noted in Figure \ref{fig:amp_dependence_1}.

To assess whether similarly good agreement could be achieved for other microchannel lengths, we fixed the amplitude at $A/W = 0.2$ and 0.4, and we varied the length across the range $1 \le L/W \le 5$ [Figure \ref{fig:Q_A}(b)]. For these geometries, $Q$ tended to decrease as $L$ decreased, qualitatively because the fluid encounters constrictions more frequently when $L$ is smaller. As before, the DPD and MPCD simulations produced almost identical results, and we found again that there was good agreement between the simulations and theoretical predictions when $L/W \gtrsim 2$. For shorter channels, however, the theoretical predictions began to deviate noticeably from the simulations. We suspected that this deviation was due to inaccuracy in the theoretical predictions, rather than the simulations, because the small parameter used in the perturbation method increases as $L/W$ decreases, potentially making the series expansion less accurate.

\begin{figure}
    \centering
    \includegraphics{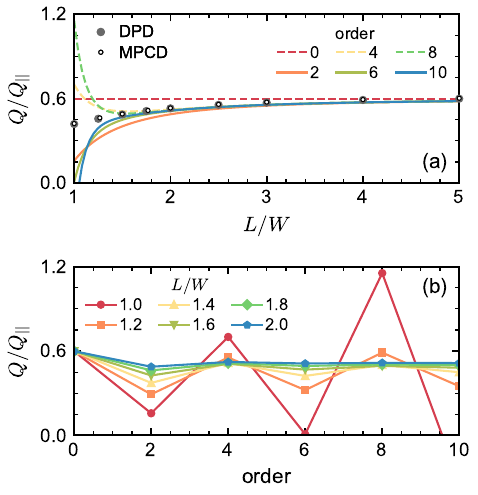}
    \caption{(a) Theoretical approximations of varying order for the volumetric flow rate $Q$ as a function of length $L$ for amplitude $A/W = 0.2$. Closed symbols are DPD simulations, open symbols are MPCD simulations, and lines are theoretical predictions. (b) Approximations for $Q$ as a function of order for different lengths $L$ and amplitude $A/W = 0.2$.}
    \label{fig:approx_degree}
\end{figure}

To test this idea, we calculated the theoretical solution for the volumetric flow rate using approximations from zeroth to tenth order as a function of length $L$ at fixed amplitude $A/W = 0.2$, and we compared these predictions to the simulation results [Figure \ref{fig:approx_degree}(a)]. Increasing the order of the approximation improved the match to the simulations at moderate lengths, but an apparent divergence emerged for small lengths; the predicted $Q$ seemed to alternate between increasingly positive and negative values for $L/W = 1$. To illustrate this behavior more clearly, we also show the theoretical predictions for $Q$ as a function of the approximation order in short microchannels ($1 \le L/W \le 2$) when $A/W = 0.2$ [Figure \ref{fig:approx_degree}(b)]. We found that $Q$ tended to converge when $L/W \ge 1.4$, was nearly oscillatory when $L/W = 1.2$, and was clearly divergent when $L/W = 1$.

\begin{figure}
    \centering
    \includegraphics{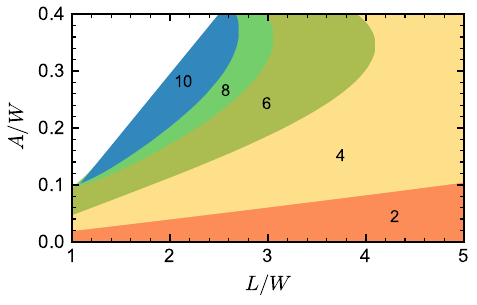}
    \caption{Microchannel lengths $L$ and amplitudes $W$ for which the series solution for the volumetric flow rate $Q$ converged to a given order. The series did not converge in the unlabeled region.}
    \label{fig:convergence}
\end{figure}

To provide guidance on use of the theoretical solution, we identified which geometries could be modeled by what order series (Figure \ref{fig:convergence}). As a simple test, we considered a series converged at order $n$ if its sum was positive for all orders of approximation and if all series terms of order $n$ and greater had absolute value less than 1\% of the tenth-order sum; otherwise, the series was considered unconverged. As expected and consistent with Figure \ref{fig:approx_degree}, the order necessary for convergence monotonically increased as the microchannel length decreased at fixed amplitude. Further, for a fixed length, microchannels with larger amplitudes typically required a higher-order series to achieve convergence. There is also a region of short lengths and large amplitudes (upper left corner) where the series did not converge even to tenth order.

\section{Conclusions}
\label{sec:Conclusions}
We investigated the flow of Newtonian fluids through sinusoidal expansion--contraction microchannels at low Reynolds numbers using complementary analytical and numerical methods. Extending prior analysis \cite{Kitanidis1997}, we first derived series solutions for the stream function and volumetric flow rate in these geometries up to tenth order using a perturbation method. We then used two popular mesoscale particle-based fluid models, DPD and MPCD, to simulate the same flows numerically. Overall, we found that the particle-based simulations and the series solution were in good agreement with each other. The fluid velocity at the expansion and contraction points not only matched between DPD and MPCD, but also agreed with the fourth-order series solution. The volumetric flow rate also showed good agreement for a range of microchannel geometries. We did note a small overestimation of the flow rates in both particle-based methods, which we attributed to the presence of some effective wall slip in these methods. 

For shorter microchannels, we found noticeable deviations between the theoretical solution and the particle-based simulations. We showed that these deviations were due to issues with convergence of the series solution. We also identified channel geometries for which the series solution for the volumetric flow rate converged for a given order of approximation. This analysis can be used as a guide for which level of approximation is needed for a specific channel geometry. Mesoscale particle-based methods like DPD or MPCD may be particularly useful for geometries where the series solution fails to converge or requires high-order terms that are cumbersome to implement. Additionally, both DPD and MPCD have straightforward ways \cite{howard_modeling_2019,DPD_overview} to incorporate various solutes, such as polymers, to simulate complex fluids using the same approach. Such extensions may enable new computational studies of viscoelastic phenomena in expansion--contraction geometries such as elastic turbulence \cite{datta_perspectives_2022}.

\section*{Supplementary Material}
See the supplementary material for alternative normalization of velocities shown in Figure \ref{fig:amp_dependence_1}, density distribution and two-dimensional velocity for MPCD fluid, parallel-plate velocity and gradients used to extract viscosities, and complete details for series solutions for stream function and volumetric flow rate up to tenth order. 

\section*{Conflicts of interest}
The authors have no conflicts to disclose.

\section*{Data Availability}
The data that support the findings of this study are available from the authors upon reasonable request.

\begin{acknowledgments}
Acknowledgment is made to the donors of the American Chemical Society Petroleum Research Fund for support of this research (Grant Nos.~65334-DNI7 to AS and 66616-DNI9 to MPH). This work used Delta at the National Center for Supercomputing Applications through allocation CHM240073 from the Advanced Cyberinfrastructure Coordination Ecosystem: Services \& Support (ACCESS) program \cite{access}, which is supported by U.S. National Science Foundation grants \#2138259, \#2138286, \#2138307, \#2137603, and \#2138296.
\end{acknowledgments}

\bibliography{references}

\end{document}


\title{Supplementary material for ``Mesoscale particle-based simulations of flow in expansion--contraction microchannels at low Reynolds number''}

\author{Tzortzis Koulaxizis}
\thanks{These authors contributed equally.}
\affiliation{Department of Chemical and Biomolecular Engineering, University of Illinois, Urbana--Champaign, Illinois 61801, USA}

\author{Clara De La Torre Garcia}
\thanks{These authors contributed equally.}
\affiliation{Department of Chemical Engineering, Auburn University, Auburn, AL 36849, USA}

\author{C. Levi Petix}
\affiliation{Department of Chemical Engineering, Auburn University, Auburn, AL 36849, USA}

\author{Antonia Statt}
\email{statt@illinois.edu}
\affiliation{Department of Chemical and Biomolecular Engineering, University of Illinois, Urbana--Champaign, Illinois 61801, USA}
\affiliation{Department of Materials Science and Engineering, The Grainger College of Engineering, University of Illinois, Urbana--Champaign, Illinois 61801, USA}

\author{Michael P. Howard}
\email{mphoward@auburn.edu}
\affiliation{Department of Chemical Engineering, Auburn University, Auburn, AL 36849, USA}

\maketitle
\section{Supplementary Figures}

\begin{figure}[!ht]
    \centering
    \includegraphics{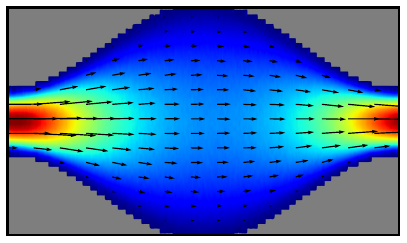}
    \caption{Same as Figure 2(b) but for the MPCD fluid.}
    \label{fig:si_y_density2}
\end{figure}

\begin{figure}[!ht]
    \centering
    \includegraphics{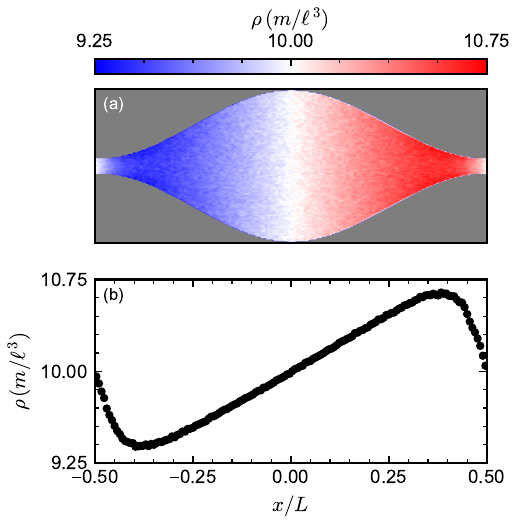}
    \caption{Average mass density $\rho$ of the MPCD fluid in the microchannel (a) as a function of $x$ and $y$ and (b) as a function of $x$ averaged over $y$.}
    \label{fig:si_x_density}
\end{figure}

\begin{figure}[!ht]
    \centering
    \includegraphics{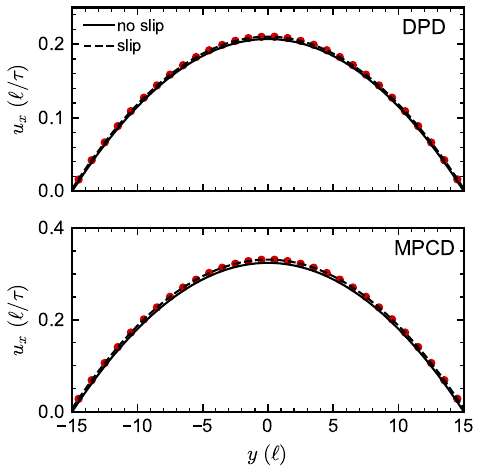}
    \caption{Velocity $u_x$ as a function of $y$ in the parallel-plate geometry for DPD (top) and MPCD (bottom). Points are simulation data. The solid line represents the theoretically expected velocity with no-slip boundary conditions and the viscosity extracted from Figure \ref{fig:si_viscosity}. The dashed lines are the same but with a best-fit slip velocity of $3.20 \times 10^{-3} \, \ell/\tau$ for DPD and $7.06 \times 10^{-3} \, \ell/\tau$ for MPCD.}
    \label{fig:si_vel}
\end{figure}

\begin{figure}[!ht]
    \centering
    \includegraphics{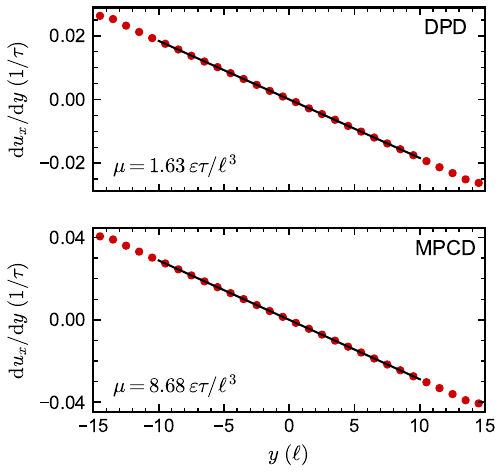}
    \caption{Gradients $\d{u_x}/\d{y}$ of velocities shown in Figure \ref{fig:si_vel}. Solid lines show the linear fits used to extract the viscosity $\mu$.}
    \label{fig:si_viscosity}
\end{figure}

\begin{figure}[!ht]
    \includegraphics{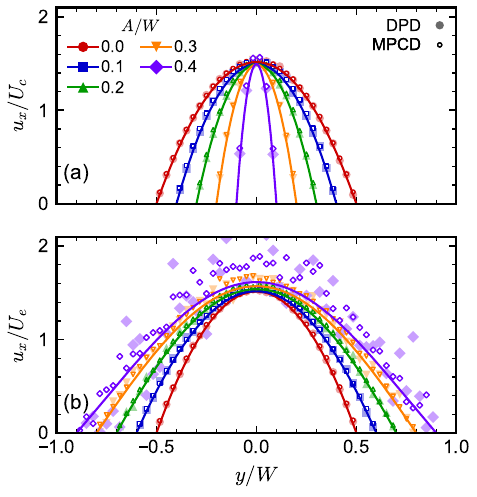}
    \caption{Same as Figure 3 but normalized using the fourth-order theoretical solution for the average velocity at the (a) contraction $U_{\rm c}$ and (b) expansion $U_{\rm e}$.}
    \label{fig:amp_dependence_2}
\end{figure}

\FloatBarrier
\section{Perturbation solution}
\label{sec:si_solution}

\subsection{Zeroth-order solution}
The partial differential equation for $n=0$ was
\begin{equation}
\partial_{0,4}\hat\psi_{0}=0,
\end{equation}
where $\partial_{i,j} \hat\psi$ is the $i$-th partial derivative with respect to $\eta$ and $j$-th partial derivative with respect to $\xi$ of $\hat\psi$. The solution for $\hat\psi_0$ was
\begin{equation}
\hat\psi_{0}=\hat f_{0} \hat{Q}_{0}
\end{equation}
with
\begin{equation}
\hat f_{0} = -\frac{1}{4} \xi \big( \xi^2 - 3 \big).
\end{equation}
The corresponding solution for $\hat Q_0$ was
\begin{equation}
\hat{Q}_{0}=\frac{1}{\hat{I}_{0}}
\end{equation}
with
\begin{equation}
\hat{I}_{0}=\frac{1}{8} \int_{-1/2}^{1/2} \hat{H}^{-3} \d{\eta}.
\end{equation}

\subsection{Second-order solution}
The partial differential equation for $n=2$ was
\begin{longalign}
\partial_{0,4}\hat\psi_{2}
&+ 2 \hat{H}_1^2 \bigg[
    6 \partial_{0,2}\hat\psi_{0}
    + \xi \Big(
        6 \partial_{0,3}\hat\psi_{0}
        + \xi \, \partial_{0,4}\hat\psi_{0}
    \Big)
\bigg]
+ 2 \hat{H}^2 \partial_{2,2}\hat\psi_{0}  \\
&= 2 \hat{H} \bigg[
    \hat{H}_2 \Big(
        2 \partial_{0,2}\hat\psi_{0}
        + \xi \, \partial_{0,3}\hat\psi_{0}
    \Big)
    + 2 \hat{H}_1 \Big(
        2 \partial_{1,2}\hat\psi_{0}
        + \xi \, \partial_{1,3}\hat\psi_{0}
    \Big)
\bigg], 
\end{longalign}
and the solution for $\hat\psi_2$ was
\begin{equation}
\hat\psi_{2} = \hat{Q}_{2} \hat f_{0} + \hat{Q}_{0} \hat f_{2}
\end{equation}
with
\begin{equation}
\hat f_{2} = \frac{3}{40} \xi \big( \xi^2 - 1 \big)^2 \big( 4 \hat{H}_1^2 - \hat{H} \hat{H}_2 \big).
\end{equation}
The corresponding solution for $\hat Q_2$ was
\begin{equation}
\hat{Q}_{2}=\frac{\hat{I}_{2}}{\hat{I}_{0}^2}
\end{equation}
with
\begin{equation}
\hat{I}_{2} = \frac{1}{20} \int_{-1/2}^{1/2} 
\big( \hat{H} \hat{H}_2 - 5 \hat{H}_1^2 \big) 
\hat{H}^{-3} \d{\eta}.
\end{equation}

\subsection{Fourth-order solution}
The partial differential equation for $n=4$ was
\begin{longalign}
\partial_{0,4}\hat\psi_{4}
&+ \hat{H}^2 \bigg[
    2 \partial_{2,2}\hat\psi_{2}
    + 3 \xi \hat{H}_2^2 \Big(
        2 \partial_{0,1}\hat\psi_{0}
        + \xi \, \partial_{0,2}\hat\psi_{0}
    \Big)
\bigg]  \\
&+ \xi \hat{H}_1^4 \bigg[
    \xi \Big(
        \xi^2 \partial_{0,4}\hat\psi_{0}
        + 12 \xi \partial_{0,3}\hat\psi_{0}
        + 36 \partial_{0,2}\hat\psi_{0}
    \Big)
    + 24 \partial_{0,1}\hat\psi_{0}
\bigg]  \\
&+ 2 \hat{H}_1^2 \bigg[
    \xi^2 \partial_{0,4}\hat\psi_{2}
    + 6 \xi \partial_{0,3}\hat\psi_{2}
    + 6 \partial_{0,2}\hat\psi_{2} \\
&\qquad- 3 \xi \hat{H} \hat{H}_2 \Big(
        \xi^2 \partial_{0,3}\hat\psi_{0}
        + 6 \xi \partial_{0,2}\hat\psi_{0}
        + 6 \partial_{0,1}\hat\psi_{0}
    \Big) \\
&\qquad+ 3 \xi \hat{H}^2 \Big(
        2 \partial_{2,1}\hat\psi_{0}
        + \xi \, \partial_{2,2}\hat\psi_{0}
    \Big)
\bigg]  \\
&+ 4 \hat{H} \hat{H}_1 \bigg(
    -2 \partial_{1,2}\hat\psi_{2}
    - \xi \, \partial_{1,3}\hat\psi_{2} \\
&\qquad+ \xi \hat{H} \Big[
        \hat{H}_3 \Big(
            2 \partial_{0,1}\hat\psi_{0}
            + \xi \, \partial_{0,2}\hat\psi_{0}
        \Big)
        + 3 \hat{H}_2 \Big(
            2 \partial_{1,1}\hat\psi_{0}
            + \xi \, \partial_{1,2}\hat\psi_{0}
        \Big)
    \Big]  \\
&\qquad - \xi \hat{H}^2 \partial_{3,1}\hat\psi_{0}
\bigg) \\
&+ \hat{H}^4 \partial_{4,0}\hat\psi_{0}  \\
&= \hat{H} \bigg[
    2 \hat{H}_2 \Big(
        2 \partial_{0,2}\hat\psi_{2}
        + \xi \, \partial_{0,3}\hat\psi_{2}
    \Big) \\
&\qquad+ 4 \xi \hat{H}_1^3 \Big(
        \xi^2 \partial_{1,3}\hat\psi_{0}
        + 6 \xi \partial_{1,2}\hat\psi_{0}
        + 6 \partial_{1,1}\hat\psi_{0}
    \Big)  \\
&\qquad + \xi \hat{H}^2 \Big(
        \hat{H}_4 \partial_{0,1}\hat\psi_{0}
        + 4 \hat{H}_3 \partial_{1,1}\hat\psi_{0}
        + 6 \hat{H}_2 \partial_{2,1}\hat\psi_{0}
    \Big)
\bigg],
\end{longalign}
and the solution for $\hat\psi_4$ was
\begin{equation}
\hat\psi_{4} = \hat{Q}_{4} \hat f_{0} + \hat{Q}_{2} \hat f_{2} + \hat{Q}_{0} \hat f_{4}
\end{equation}
with
\begin{longalign}
\hat f_{4} = &-\frac{1}{5600}
    \xi \big( \xi^2 - 1 \big)^2 \Big[
        24 \left( 75 \xi^2 + 17 \right) \hat{H}_1^4 \\
&+ \hat{H}^2 \Big[
            \left( 19 - 15 \xi^2 \right) \hat{H} \hat{H}_4
            + 90 \left( 2 \xi^2 - 3 \right) \hat{H}_2^2
        \Big] \\
&+ 8 \left( 30 \xi^2 - 31 \right) \hat{H}^2 \hat{H}_3 \hat{H}_1
        - 36 \left( 50 \xi^2 - 19 \right) \hat{H} \hat{H}_1^2 \hat{H}_2
    \Big].
\end{longalign}
The corresponding solution for $\hat Q_4$ was
\begin{equation}
\hat{Q}_{4}=\frac{\hat{I}_{2}^2-\hat{I}_{0} \hat{I}_{4}}{\hat{I}_{0}^3}
\end{equation}
with
\begin{longalign}
\hat{I}_{4} = \frac{1}{1400} & \int_{-1/2}^{1/2}
\big(
    -4 \hat{H}^3 \hat{H}_{4}
    + 54 \hat{H}^2 \hat{H}_2^2
    + 87 \hat{H}_1^4 \\
    &+ 56 \hat{H}^2 \hat{H}_{3} \hat{H}_1
    - 306 \hat{H} \hat{H}_1^2 \hat{H}_2
\big) 
\hat{H}^{-3} \d{\eta}.
\end{longalign}

\subsection{Sixth-order solution}
The partial differential equation for $n=6$ was
\begin{longalign}
\partial_{0,4}\hat\psi_{6}
&+ \hat{H}^2 \bigg[
    2 \partial_{2,2}\hat\psi_{4}
    + 3 \xi \hat{H}_2^2 \Big(
        2 \partial_{0,1}\hat\psi_{2}
        + \xi \, \partial_{0,2}\hat\psi_{2}
    \Big)
\bigg] \\
&+ \xi \hat{H}_1^4 \bigg[
    \xi \Big(
        \xi^2 \partial_{0,4}\hat\psi_{2}
        + 12 \xi \partial_{0,3}\hat\psi_{2}
        + 36 \partial_{0,2}\hat\psi_{2}
    \Big)
    + 24 \partial_{0,1}\hat\psi_{2}
\bigg] \\
&+ 2 \hat{H}_1^2 \bigg[
    \xi^2 \partial_{0,4}\hat\psi_{4}
    + 6 \xi \partial_{0,3}\hat\psi_{4}
    + 6 \partial_{0,2}\hat\psi_{4} \\
&\qquad- 3 \xi \hat{H} \hat{H}_2 \Big(
        \xi^2 \partial_{0,3}\hat\psi_{2}
        + 6 \xi \partial_{0,2}\hat\psi_{2}
        + 6 \partial_{0,1}\hat\psi_{2}
    \Big) \\
&\qquad
    + 3 \xi \hat{H}^2 \Big(
        2 \partial_{2,1}\hat\psi_{2}
        + \xi \, \partial_{2,2}\hat\psi_{2}
    \Big)
\bigg] \\
&+ 4 \hat{H} \hat{H}_1 \bigg(
    -2 \partial_{1,2}\hat\psi_{4}
    - \xi \partial_{1,3}\hat\psi_{4} \\
&\qquad+ \xi \hat{H} \Big[
        \hat{H}_3 \Big(
            2 \partial_{0,1}\hat\psi_{2}
            + \xi \partial_{0,2}\hat\psi_{2}
        \Big)
        + 3 \hat{H}_2 \Big(
            2 \partial_{1,1}\hat\psi_{2}
            + \xi \partial_{1,2}\hat\psi_{2}
        \Big)
    \Big] \\
&\qquad
    - \xi \hat{H}^2 \partial_{3,1}\hat\psi_{2}
\bigg) \\
&+ \hat{H}^4 \partial_{4,0}\hat\psi_{2} \\
&= \hat{H} \bigg[
    2 \hat{H}_2 \Big(
        2 \partial_{0,2}\hat\psi_{4}
        + \xi \partial_{0,3}\hat\psi_{4}
    \Big) \\
&\qquad+ 4 \xi \hat{H}_1^3 \Big(
        \xi^2 \partial_{1,3}\hat\psi_{2}
        + 6 \xi \partial_{1,2}\hat\psi_{2}
        + 6 \partial_{1,1}\hat\psi_{2}
    \Big) \\
    &\qquad
    + \xi \hat{H}^2 \Big(
        \hat{H}_4 \partial_{0,1}\hat\psi_{2}
        + 4 \hat{H}_3 \partial_{1,1}\hat\psi_{2}
        + 6 \hat{H}_2 \partial_{2,1}\hat\psi_{2}
    \Big)
\bigg],
\end{longalign}
and the solution for $\hat\psi_6$ was
\begin{equation}
\hat\psi_{6}= \hat{Q}_{6} \hat f_{0} + \hat{Q}_{4} \hat f_{2} + \hat{Q}_{2} \hat f_{4} + \hat{Q}_{0} \hat f_{6}
\end{equation}
with
\begin{longalign}
\hat f_{6} = &\frac{1}{504000}
    \xi \big( \xi^2 - 1 \big)^2 \Big(
        96 \big( 1750 \xi^4 + 575 \xi^2 + 192 \big) \hat{H}_1^6 \\
        &+ 480 \big( 125 \xi^4 - 86 \xi^2 - 42 \big) \hat{H}^2 \hat{H}_3 \hat{H}_1^3 - 72 \big( 4375 \xi^4 - 550 \xi^2 - 1248 \big) \hat{H} \hat{H}_1^4 \hat{H}_2 \\
        &- 12 \hat{H}^2 \hat{H}_1^2 \Big[
            5 \big( 125 \xi^4 - 173 \xi^2 + 36 \big) \hat{H} \hat{H}_4 - 18 \big( 625 \xi^4 - 520 \xi^2 - 248 \big) \hat{H}_2^2
        \Big] \\
    &+ \hat{H}^3 \Big[
            -12 \big( 625 \xi^4 - 1270 \xi^2 + 867 \big) \hat{H}_2^3
            + \hat{H} \Big[
                \big( -25 \xi^4 + 70 \xi^2 + 3 \big) \hat{H} \hat{H}_6 \\
    &\qquad
                + 8 \big( 125 \xi^4 - 455 \xi^2 + 468 \big) \hat{H}_3^2
            \Big]
            + 12 \big( 125 \xi^4 - 410 \xi^2 + 399 \big) \hat{H} \hat{H}_4 \hat{H}_2
        \Big] \\
    &+ 24 \hat{H}^3 \hat{H}_1 \Big[
            \big( 25 \xi^4 - 55 \xi^2 + 36 \big) \hat{H} \hat{H}_5
            - 2 \big( 625 \xi^4 - 1135 \xi^2 + 468 \big) \hat{H}_3 \hat{H}_2
        \Big]
    \Big).
\end{longalign}
The corresponding solution for $\hat Q_6$ was
\begin{equation}
\hat{Q}_{6}=\frac{\hat{I}_{2}^3-2 \hat{I}_{0} \hat{I}_{4} \hat{I}_{2}+\hat{I}_{0}^2 \hat{I}_{6}}{\hat{I}_{0}^4}
\end{equation}
with
\begin{longalign}
\hat{I}_{6} = \frac{1}{10500} & \int_{-1/2}^{1/2}
\bigg[
    128 \hat{H}_1^6
    - 136 \hat{H}^2 \hat{H}_3 \hat{H}_1^3
    - 1176 \hat{H} \hat{H}_1^4 \hat{H}_2 \\
&+ 4 \hat{H}^2 \hat{H}_1^2 \Big(
        23 \hat{H} \hat{H}_4
        - 108 \hat{H}_2^2
    \Big) \\
&+ \hat{H}^3 \Big(
        207 \hat{H}_2^3
        + \hat{H} \big(
            \hat{H} \hat{H}_6
            - 58 \hat{H}_3^2
        \big)
        - 76 \hat{H} \hat{H}_4 \hat{H}_2
    \Big) \\
&+ 2 \hat{H}^3 \hat{H}_1 \Big(
        300 \hat{H}_3 \hat{H}_2
        - 7 \hat{H} \hat{H}_5
    \Big)
\bigg] \hat{H}^{-3} \d{\eta}.
\end{longalign}
\subsection{Eighth-order solution}
The partial differential equation for $n=8$ was
\begin{longalign}
\partial_{0,4}\hat\psi_{8}
&+ \hat{H}^2 \bigg[
    2 \partial_{2,2}\hat\psi_{6}
    + 3 \xi \hat{H}_2^2 \Big(
        2 \partial_{0,1}\hat\psi_{4}
        + \xi \, \partial_{0,2}\hat\psi_{4}
    \Big)
\bigg] \\
&+ \xi \hat{H}_1^4 \bigg[
    \xi \Big(
        \xi^2 \partial_{0,4}\hat\psi_{4}
        + 12 \xi \partial_{0,3}\hat\psi_{4}
        + 36 \partial_{0,2}\hat\psi_{4}
    \Big)
    + 24 \partial_{0,1}\hat\psi_{4}
\bigg] \\
&+ 2 \hat{H}_1^2 \bigg[
    \xi^2 \partial_{0,4}\hat\psi_{6}
    + 6 \xi \partial_{0,3}\hat\psi_{6}
    + 6 \partial_{0,2}\hat\psi_{6} \\
&\qquad- 3 \xi \hat{H} \hat{H}_2 \Big(
        \xi^2 \partial_{0,3}\hat\psi_{4}
        + 6 \xi \partial_{0,2}\hat\psi_{4}
        + 6 \partial_{0,1}\hat\psi_{4}
    \Big) \\
&\qquad
    + 3 \xi \hat{H}^2 \Big(
        2 \partial_{2,1}\hat\psi_{4}
        + \xi \, \partial_{2,2}\hat\psi_{4}
    \Big)
 \bigg] \\
&+ 4 \hat{H} \hat{H}_1 \bigg(
    -2 \partial_{1,2}\hat\psi_{6}
    - \xi \partial_{1,3}\hat\psi_{6}
    + \xi \hat{H} \Big[
        \hat{H}_3 \Big(
            2 \partial_{0,1}\hat\psi_{4}
            + \xi \partial_{0,2}\hat\psi_{4}
        \Big) \\
&\qquad
        + 3 \hat{H}_2 \Big(
            2 \partial_{1,1}\hat\psi_{4}
            + \xi \partial_{1,2}\hat\psi_{4}
        \Big)
    \Big]
    - \xi \hat{H}^2 \partial_{3,1}\hat\psi_{4}
\bigg) \\
&+ \hat{H}^4 \partial_{4,0}\hat\psi_{4} \\
&= \hat{H} \bigg[
    2 \hat{H}_2 \Big(
        2 \partial_{0,2}\hat\psi_{6}
        + \xi \partial_{0,3}\hat\psi_{6}
    \Big) \\
&\qquad+ 4 \xi \hat{H}_1^3 \Big(
        \xi^2 \partial_{1,3}\hat\psi_{4}
        + 6 \xi \partial_{1,2}\hat\psi_{4}
        + 6 \partial_{1,1}\hat\psi_{4}
    \Big) \\
 &\qquad
    + \xi \hat{H}^2 \Big(
        \hat{H}_4 \partial_{0,1}\hat\psi_{4}
        + 4 \hat{H}_3 \partial_{1,1}\hat\psi_{4}
        + 6 \hat{H}_2 \partial_{2,1}\hat\psi_{4}
    \Big)
\bigg],
\end{longalign}
and the solution for $\hat\psi_8$ was
\begin{equation}
\hat\psi_{8}= \hat{Q}_{8} \hat f_{0} + \hat{Q}_{6} \hat f_{2} + \hat{Q}_{4} \hat f_{4} + \hat{Q}_{2} \hat f_{6} + \hat{Q}_{0} \hat f_{8}
\end{equation}
with
\begin{longalign}
\hat f_{8} = &-\frac{1}{1552320000}
\xi \left(\xi^{2} - 1\right)^2 \\
&\times\Big(
    1152 \left(459375 \xi^{6} + 177625 \xi^{4} + 84800 \xi^{2} + 31168\right) \hat{H}_1^{8} \\
&
    + 384 \left(857500 \xi^{6} - 335125 \xi^{4} - 423240 \xi^{2} - 112572\right) \hat{H}^2 \hat{H}_3 \hat{H}_1^{5} \\
&
    - 576 \left(2572500 \xi^{6} + 91875 \xi^{4} - 542400 \xi^{2} - 560752\right) \hat{H} \hat{H}_1^{6} \hat{H}_2 \\
&
    - 48 \hat{H}^2 \hat{H}_1^{4} \Big[
        \left(1071875 \xi^{6} - 984375 \xi^{4} - 559520 \xi^{2} + 418272\right) \hat{H} \hat{H}_4 \\
&\qquad
       - 42 \left(612500 \xi^{6} - 280625 \xi^{4} - 358320 \xi^{2} - 83256\right) \hat{H}_2^{2}
    \Big] \\
&
    + 192 \hat{H}^3 \hat{H}_1^{3} \Big[
        5 \left(6125 \xi^{6} - 9695 \xi^{4} - 72 \xi^{2} + 5316\right) \hat{H} \hat{H}_5 \\
&\qquad
        - 2 \left(1071875 \xi^{6} - 1176875 \xi^{4} - 660060 \xi^{2} + 731706\right) \hat{H}_3 \hat{H}_2
    \Big] \\
&
    - 8 \hat{H}^3 \hat{H}_1^{2} \Big[
        36 \left(1071875 \xi^{6} - 1273125 \xi^{4} - 719735 \xi^{2} + 1054259\right) \hat{H}_2^{3} \\
&\qquad
        + \hat{H} \Big(
            \left(61250 \xi^{6} - 148925 \xi^{4} + 96960 \xi^{2} - 16437\right) \hat{H} \hat{H}_6 \\
&\qquad\qquad
            - 168 \left(21875 \xi^{6} - 45625 \xi^{4} + 6720 \xi^{2} + 30784\right) \hat{H}_3^{2}
        \Big) \\
&\qquad
        - 36 \left(153125 \xi^{6} - 300125 \xi^{4} + 34995 \xi^{2} + 195787\right) \hat{H} \hat{H}_4 \hat{H}_2
    \Big] \\
&
    + \hat{H}^4 \Big[
        72 \left(153125 \xi^{6} - 357875 \xi^{4} + 98685 \xi^{2} + 196733\right) \hat{H}_2^{4} \\
&\qquad
        - 12 \left(306250 \xi^{6} - 1148875 \xi^{4} + 1814040 \xi^{2} - 2077431\right) \hat{H} \hat{H}_4 \hat{H}_2^{2} \\
&\qquad
        + 8 \hat{H} \Big(
            \left(12250 \xi^{6} - 62125 \xi^{4} + 128820 \xi^{2} - 91569\right) \hat{H} \hat{H}_6 \\
&\qquad\qquad
+ \left(-612500 \xi^{6} + 2413250 \xi^{4} - 4164000 \xi^{2} + 4989234\right) \hat{H}_3^{2}
        \Big) \hat{H}_2 \\
&\qquad
        + \hat{H}^2 \Big(
            \left(-875 \xi^{6} + 4025 \xi^{4} + 2655 \xi^{2} - 22893\right) \hat{H} \hat{H}_8 \\
&\qquad\qquad
            + 2 \left(61250 \xi^{6} - 387625 \xi^{4} + 911180 \xi^{2} - 476853\right) \hat{H}_4^{2} \\
&\qquad\qquad
            + 16 \left(12250 \xi^{6} - 73675 \xi^{4} + 169080 \xi^{2} - 92823\right) \hat{H}_3 \hat{H}_5
        \Big)
    \Big] \\
&+ 16 \hat{H}^4 \hat{H}_1 \Big[
        3 \left(-61250 \xi^{6} + 195125 \xi^{4} - 201240 \xi^{2} + 140517\right) \hat{H} \hat{H}_5 \hat{H}_2 \\
&\qquad
        + 180 \left(30625 \xi^{6} - 67725 \xi^{4} + 15909 \xi^{2} + 46259\right) \hat{H}_3 \hat{H}_2^{2} \\
&\qquad
        + \hat{H} \Big(
            \left(1750 \xi^{6} - 6125 \xi^{4} + 8440 \xi^{2} - 16881\right) \hat{H} \hat{H}_7 \\
&\qquad\qquad
            - 35 \left(8750 \xi^{6} - 31175 \xi^{4} + 36084 \xi^{2} - 27579\right) \hat{H}_3 \hat{H}_4
        \Big)
    \Big]
\Big).
\end{longalign}
The corresponding solution for $\hat Q_8$ was
\begin{equation}
\hat{Q}_{8}=\frac{\hat{I}_{2}^4-3 \hat{I}_{0} \hat{I}_{4} \hat{I}_{2}^2+2 \hat{I}_{0}^2 \hat{I}_{6} \hat{I}_{2}+\hat{I}_{0}^2 \hat{I}_{4}^2-\hat{I}_{0}^3 \hat{I}_{8}}{\hat{I}_{0}^5}
\end{equation}
with
\begin{longalign}
\hat{I}_{8} = \frac{1}{72765000} &\int_{-1/2}^{1/2}
\bigg[
    398016 \hat{H}_1^{8}
    - 2408400 \hat{H}^2 \hat{H}_3 \hat{H}_1^{5}
    - 6462576 \hat{H} \hat{H}_1^{6} \hat{H}_2 \\
&
    + 144 \hat{H}^2 \hat{H}_1^{4} \Big(
        5386 \hat{H} \hat{H}_4
        - 84657 \hat{H}_2^{2}
    \Big) \\
&+ 36 \hat{H}^3 \hat{H}_1^{3} \Big(
        248032 \hat{H}_3 \hat{H}_2
        - 1059 \hat{H} \hat{H}_5
    \Big) \\
&
    - 6 \hat{H}^3 \hat{H}_1^{2} \Big[
        -1408779 \hat{H}_2^{3}
        + \hat{H} \Big(
            1891 \hat{H} \hat{H}_6
            + 46202 \hat{H}_3^{2}
        \Big)\\
&\qquad+ 69708 \hat{H} \hat{H}_4 \hat{H}_2
    \Big] \\
&
    + \hat{H}^4 \Big[
        -101169 \hat{H}_2^{4}
        - 598284 \hat{H} \hat{H}_4 \hat{H}_2^{2} \\
&\qquad+ 4 \hat{H} \Big(
            7147 \hat{H} \hat{H}_6
            - 238641 \hat{H}_3^{2}
        \Big) \hat{H}_2 \\
&\qquad
        + 2 \hat{H}^2 \Big(
            62 \hat{H} \hat{H}_8
            + 26827 \hat{H}_4^{2}
            + 39028 \hat{H}_3 \hat{H}_5
        \Big)
    \Big] \\
&
    + 4 \hat{H}^4 \hat{H}_1 \Big(
        770 \hat{H}_7 \hat{H}^2
        - 36621 \hat{H}_5 \hat{H} \hat{H}_2\\
&\qquad- 79746 \hat{H}_3 \hat{H}_4 \hat{H}
        - 220014 \hat{H}_3 \hat{H}_2^{2}
    \Big)
\bigg] \hat{H}^{-3} \d{\eta}.
\end{longalign}
        \subsection{Tenth-order solution}
The partial differential equation for $n=10$ was
\begin{longalign}
\partial_{0,4}\hat{\psi}_{10}
&+ \hat{H}^2 \Big[
    2 \partial_{2,2}\hat{\psi}_{8}
    + 3 \xi \hat{H}_2^2 \big(
        2 \partial_{0,1}\hat{\psi}_{6}
        + \xi \partial_{0,2}\hat{\psi}_{6}
    \big)
\Big] \\
& + \xi \hat{H}_1^4 \Big[
    \xi \big(
        \xi^2 \partial_{0,4}\hat{\psi}_{6}
        + 12 \xi \partial_{0,3}\hat{\psi}_{6}
        + 36 \partial_{0,2}\hat{\psi}_{6}
    \big)
    + 24 \partial_{0,1}\hat{\psi}_{6}
\Big] \\
& + 2 \hat{H}_1^2 \Big[
    \xi^2 \partial_{0,4}\hat{\psi}_{8}
    + 6 \xi \partial_{0,3}\hat{\psi}_{8}
    + 6 \partial_{0,2}\hat{\psi}_{8} \\
&\qquad- 3 \xi \hat{H} \hat{H}_2 \big(
        \xi^2 \partial_{0,3}\hat{\psi}_{6}
        + 6 \xi \partial_{0,2}\hat{\psi}_{6}
        + 6 \partial_{0,1}\hat{\psi}_{6}
    \big) \\
&\qquad
    + 3 \xi \hat{H}^2 \big(
        2 \partial_{2,1}\hat{\psi}_{6}
        + \xi \partial_{2,2}\hat{\psi}_{6}
    \big)
    \Big] \\
& + 4 \hat{H} \hat{H}_1 \Big[
    - 2 \partial_{1,2}\hat{\psi}_{8}
    - \xi \partial_{1,3}\hat{\psi}_{8} \\
&\qquad+ \xi \hat{H} \big[
        \hat{H}_3 \big(
            2 \partial_{0,1}\hat{\psi}_{6}
            + \xi \partial_{0,2}\hat{\psi}_{6}
        \big)
        + 3 \hat{H}_2 \big(
            2 \partial_{1,1}\hat{\psi}_{6}
            + \xi \partial_{1,2}\hat{\psi}_{6}
        \big)
    \big] \\
& \qquad
    - \xi \hat{H}^2 \partial_{3,1}\hat{\psi}_{6}
\Big] \\
&+ \hat{H}^4 \partial_{4,0}\hat{\psi}_{6} \\
&= \hat{H} \Big[
    2 \hat{H}_2 \big(
        2 \partial_{0,2}\hat{\psi}_{8}
        + \xi \partial_{0,3}\hat{\psi}_{8}
    \big) \\
&\qquad+ 4 \xi \hat{H}_1^3 \big(
        \xi^2 \partial_{1,3}\hat{\psi}_{6}
        + 6 \xi \partial_{1,2}\hat{\psi}_{6}
        + 6 \partial_{1,1}\hat{\psi}_{6}
    \big) \\
&\qquad
    + \xi \hat{H}^2 \big(
        \hat{H}_4 \partial_{0,1}\hat{\psi}_{6}
        + 4 \hat{H}_3 \partial_{1,1}\hat{\psi}_{6}
        + 6 \hat{H}_2 \partial_{2,1}\hat{\psi}_{6}
    \big)
\Big].
\end{longalign}
and the solution for $\hat\psi_{10}$ was
\begin{equation}
\hat\psi_{10} = \hat{Q}_{10} \hat f_{0} + \hat{Q}_{8} \hat f_{2} + \hat{Q}_{6} \hat f_{4} + \hat{Q}_{4} \hat f_{6} + \hat{Q}_{2} \hat f_{8} + \hat{Q}_{0} \hat f_{10}
\end{equation}
with
\begin{longalign}
\hat f_{10} = &\frac{1}{605404800000}
\xi \left(\xi^{2} - 1\right)^2 \\
&\times \Big(
    13824 \left(15159375 \xi^{8} + 6431250 \xi^{6} + 3584000 \xi^{4} + 1873600 \xi^{2} + 718976\right) \hat{H}_1^{10} \\
    &
    - 3456 \left(227390625 \xi^{8} + 31972500 \xi^{6} - 26684000 \xi^{4} - 47202400 \xi^{2} - 39851712\right) \\
    &\qquad\times\hat{H} \hat{H}_2 \hat{H}_1^{8} \\
    &
    + 9216 \left(20671875 \xi^{8} - 4042500 \xi^{6} - 8879875 \xi^{4} - 7016270 \xi^{2} - 875717\right) \\
    &\qquad\times \hat{H}^2 \hat{H}_3 \hat{H}_1^{7} \\
    &
    - 1152 \hat{H}^2 \Big[\\
    &\qquad6 \left(-144703125 \xi^{8} + 33871250 \xi^{6} + 71966000 \xi^{4} + 57658400 \xi^{2} + 4216848\right) \hat{H}_2^{2} \\
        &\qquad
        + \left(28940625 \xi^{8} - 17364375 \xi^{6} - 18642250 \xi^{4} - 4761410 \xi^{2} + 11257462\right) \hat{H} \hat{H}_4 \\
    &\qquad \Big]\hat{H}_1^{6} \\
    &
    + 2304 \hat{H}^3 \Big[
        \left(1929375 \xi^{8} - 2113125 \xi^{6} - 1261450 \xi^{4} + 857920 \xi^{2} + 1209186\right) \hat{H} \hat{H}_5 \\
        &\qquad
        - 6 \left(28940625 \xi^{8} - 20151250 \xi^{6} - 22106750 \xi^{4} - 4527735 \xi^{2} + 17660196\right) \hat{H}_2 \hat{H}_3 \\
    &\qquad \Big] \hat{H}_1^{5} \\
    &
    - 48 \hat{H}^3 \Big[ \\
        &\qquad 36 \left(289406250 \xi^{8} - 215446875 \xi^{6} - 242054375 \xi^{4} - 40498225 \xi^{2} + 237892709\right) \hat{H}_2^{3} \\
        &\qquad
        - 12 \left(96468750 \xi^{8} - 123571875 \xi^{6} - 73253125 \xi^{4} + 73474795 \xi^{2} + 77895771\right) \\
        &\qquad\qquad \times\hat{H} \hat{H}_4 \hat{H}_2 \\
        &\qquad
        + \hat{H} \Big(
            \left(9646875 \xi^{8} - 16537500 \xi^{6} - 2378000 \xi^{4} + 14844560 \xi^{2} - 3137439\right) \hat{H} \hat{H}_6 \\
            &\qquad\quad\quad
            - 16 \left(48234375 \xi^{8} - 64771875 \xi^{6} - 38335250 \xi^{4} \right.\\
            &\qquad\qquad\qquad\left.+ 41775710 \xi^{2} + 41678022\right) \hat{H}_3^{2}
        \Big) \\
    &\qquad\Big] \hat{H}_1^{4} \\
    &
    + 192 \hat{H}^4 \Big[ \\
        &\qquad 60 \left(19293750 \xi^{8} - 27103125 \xi^{6} - 16317875 \xi^{4} + 21635945 \xi^{2} + 16725957\right) \\
        &\qquad\qquad \times\hat{H}_3 \hat{H}_2^{2} \\
        &\qquad
        + 3 \left(-9646875 \xi^{8} + 19722500 \xi^{6} + 1269100 \xi^{4} - 23953920 \xi^{2} + 15839739\right) \\
        &\qquad\qquad \times \hat{H} \hat{H}_5 \hat{H}_2 \\
        &\qquad
        + \hat{H} \Big(
            3 \left(65625 \xi^{8} - 164500 \xi^{6} + 66550 \xi^{4} + 168900 \xi^{2} - 326399\right) \hat{H} \hat{H}_7 \\
            &\qquad\qquad
            - 5 \left(9646875 \xi^{8} - 21315000 \xi^{6} - 728950 \xi^{4} \right.\\
            &\qquad\qquad\qquad\left.+ 28438816 \xi^{2} - 21337101\right) \hat{H}_3 \hat{H}_4
        \Big) \\
    &\qquad \Big] \hat{H}_1^{3} \\
    &
    - 12 \hat{H}^4 \Big[ \\
        &\qquad -72 \left(96468750 \xi^{8} - 141487500 \xi^{6} - 87956125 \xi^{4}\right. \\
        &\qquad\qquad\left.+ 135287650 \xi^{2} + 72200289\right) \hat{H}_2^{4} \\
        &\qquad
        + 36 \left(48234375 \xi^{8} - 110556250 \xi^{6} - 238750 \xi^{4} \right. \\
        &\qquad\qquad\left.+ 177654830 \xi^{2} - 200028157\right) \hat{H} \hat{H}_4 \hat{H}_2^{2} \\
        &\qquad
        - 8 \hat{H} \Big( \\
            &\qquad\qquad\left(4134375 \xi^{8} - 12752250 \xi^{6} + 8138150 \xi^{4} \right. \\
            &\qquad\qquad\qquad\left.+ 14923030 \xi^{2} - 43155369\right) \hat{H} \hat{H}_6 \\
            &\qquad\qquad
            - 6 \left(48234375 \xi^{8} - 114537500 \xi^{6} + 1773000 \xi^{4} \right.\\
            &\qquad\qquad\qquad\left.+ 196787840 \xi^{2} - 235828467\right) \hat{H}_3^{2}
        \Big) \hat{H}_2 \\
        &\qquad\
        + \hat{H}^2 \Big( \\
         &\qquad\qquad-6 \left(6890625 \xi^{8} - 24438750 \xi^{6} + 18752750 \xi^{4} \right. \\
            &\qquad\qquad\qquad\left.+ 30232570 \xi^{2} - 95120683\right) \hat{H}_4^{2} \\
            &\qquad\qquad
            - 80 \left(826875 \xi^{8} - 2837100 \xi^{6} + 2095370 \xi^{4} \right.\\
            &\qquad\qquad\qquad\left.+ 3470572 \xi^{2} - 10727589\right) \hat{H}_3 \hat{H}_5 \\
            &\qquad\qquad
            + 3 \left(65625 \xi^{8} - 232750 \xi^{6} + 266100 \xi^{4} \right. \\
            &\qquad\qquad\qquad\left.+ 95710 \xi^{2} - 1625853\right) \hat{H} \hat{H}_8
        \Big) \\
    &\qquad\qquad\Big] \hat{H}_1^{2} \\
    &
    + 8 \hat{H}^5 \Big[ \\
        &\qquad-72 \left(48234375 \xi^{8} - 118518750 \xi^{6} + 2283250 \xi^{4} \right. \\
        &\qquad\qquad\left.+ 243720050 \xi^{2} - 364658077\right) \hat{H}_3 \hat{H}_2^{3} \\
        &\qquad
        + 36 \left(4134375 \xi^{8} - 14663250 \xi^{6} + 12953350 \xi^{4} \right.\\
        &\qquad\qquad\left.+ 16786190 \xi^{2} - 42792513\right) \hat{H} \hat{H}_5 \hat{H}_2^{2} \\
        &\qquad
        - 12 \hat{H} \Big[
            \left(196875 \xi^{8} - 903000 \xi^{6} + 1532800 \xi^{4} - 1891540 \xi^{2} + 3981057\right) \hat{H} \hat{H}_7 \\
            &\qquad\qquad
            - 6 \left(6890625 \xi^{8} - 26031250 \xi^{6} + 25577750 \xi^{4} \right. \\
            &\qquad\qquad\qquad\left.+ 29344670 \xi^{2} - 72913563\right) \hat{H}_3 \hat{H}_4
        \Big] \hat{H}_2 \\
        &\qquad
        + \hat{H} \Big( \\
            &\qquad\qquad80 \left(1378125 \xi^{8} - 5365500 \xi^{6} + 5655050 \xi^{4} \right. \\
            &\qquad\qquad\qquad\left.+ 5880244 \xi^{2} - 13913847\right) \hat{H}_3^{3} \\
            &\qquad\qquad
            - 12 \left(459375 \xi^{8} - 2425500 \xi^{6} + 4703200 \xi^{4} \right.\\
            &\qquad\qquad\qquad\left.- 7378720 \xi^{2} + 18299277\right) \hat{H} \hat{H}_6 \hat{H}_3 \\
            &\qquad\qquad
            + \hat{H} \Big[
                5 \left(2625 \xi^{8} - 12950 \xi^{6} + 27960 \xi^{4} - 181666 \xi^{2} + 416671\right) \hat{H} \hat{H}_9 \\
                &\qquad\qquad\qquad
                - 6 \left(1378125 \xi^{8} - 7754250 \xi^{6} + 15785300 \xi^{4} \right.\\
                &\qquad\qquad\qquad\qquad\left.- 26243510 \xi^{2} + 66543231\right) \hat{H}_4 \hat{H}_5
            \Big]
        \Big) \\
    &\qquad\qquad\Big] \hat{H}_1 \\
    &
    - \hat{H}^5 \Big[ \\
        &\qquad 216 \left(9646875 \xi^{8} - 24500000 \xi^{6} + 269125 \xi^{4} + 61417450 \xi^{2} - 100483194\right) \hat{H}_2^{5} \\
        &\qquad
        - 720 \left(1378125 \xi^{8} - 5365500 \xi^{6} + 5651475 \xi^{4} + 3873954 \xi^{2} - 2215966\right) \hat{H} \hat{H}_4 \hat{H}_2^{3} \\
        &\qquad
        + 72 \hat{H} \Big(
            5 \left(91875 \xi^{8} - 501025 \xi^{6} + 1234115 \xi^{4} - 2835687 \xi^{2} + 4460866\right) \hat{H} \hat{H}_6 \\
            &\qquad\qquad
            - 4 \left(6890625 \xi^{8} - 27623750 \xi^{6} + 30472250 \xi^{4} \right. \\
            &\qquad\qquad\qquad\left.+ 16393290 \xi^{2} + 5007657\right) \hat{H}_3^{2}
        \Big) \hat{H}_2^{2} \\
        &\qquad
        - 12 \hat{H}^2 \Big( \\
            &\qquad\qquad-5 \left(1378125 \xi^{8} - 8470875 \xi^{6} + 24694525 \xi^{4} - 64110781 \xi^{2} + 91602798\right) \hat{H}_4^{2} \\
            &\qquad\qquad
            - 8 \left(1378125 \xi^{8} - 8232000 \xi^{6} + 23152400 \xi^{4} \right. \\
            &\qquad\qquad\qquad\left.- 58865060 \xi^{2} + 85653543\right) \hat{H}_3 \hat{H}_5 \\
            &\qquad\qquad
            + \left(39375 \xi^{8} - 273875 \xi^{6} + 900075 \xi^{4} - 1980025 \xi^{2} + 1332594\right) \hat{H} \hat{H}_8
        \Big) \hat{H}_2 \\
        &\qquad
        + \hat{H}^2 \Big( \\
            &\qquad\qquad 16 \left(6890625 \xi^{8} - 43548750 \xi^{6} + 138297500 \xi^{4} \right. \\
            &\qquad\qquad\qquad\left.- 381048170 \xi^{2} + 499790907\right) \hat{H}_4 \hat{H}_3^{2} \\
            &\qquad\qquad
            - 32 \left(39375 \xi^{8} - 330750 \xi^{6} + 1233200 \xi^{4} \right.\\
            &\qquad\qquad\qquad\left.- 1698770 \xi^{2} - 777423\right) \hat{H} \hat{H}_7 \hat{H}_3 \\
            &\qquad\qquad
            + \hat{H} \Big(
                -120 \left(11025 \xi^{8} - 105350 \xi^{6} + 418200 \xi^{4} - 448210 \xi^{2} - 589233\right) \hat{H}_5^{2} \\
                &\qquad\qquad\qquad
                - 200 \left(11025 \xi^{8} - 102165 \xi^{6} + 400117 \xi^{4} \right.\\
                &\qquad\qquad\qquad\qquad\left.- 452383 \xi^{2} - 500274\right) \hat{H}_4 \hat{H}_6 \\
                &\qquad\qquad\qquad
                + \left(2625 \xi^{8} - 17500 \xi^{6} - 34050 \xi^{4} + 489940 \xi^{2} - 845239\right) \hat{H} \hat{H}_{10}
            \Big) \\
&\qquad\qquad \Big] \\
&\qquad\Big).
\end{longalign}
The corresponding solution for $\hat Q_{10}$ was
\begin{equation}
\hat{Q}_{10}=\frac{\hat{I}_{2}^5-4 \hat{I}_{0} \hat{I}_{4} \hat{I}_{2}^3+3 \hat{I}_{0}^2 \hat{I}_{6} \hat{I}_{2}^2+3 \hat{I}_{0}^2 \hat{I}_{4}^2 \hat{I}_{2}-2 \hat{I}_{0}^3 \hat{I}_{8} \hat{I}_{2}-2 \hat{I}_{0}^3 \hat{I}_{4} \hat{I}_{6}+\hat{I}_{0}^4 \hat{I}_{10}}{\hat{I}_{0}^6}
\end{equation}
with
\begin{longalign}
\hat{I}_{10} = \frac{1}{2364862500} &\int_{-1/2}^{1/2} \Big[
 \, 7649856 \hat{H}_1^{10}
- 111217608 \hat{H}^2 \hat{H}_{3} \hat{H}_1^7
- 184424472 \hat{H} \hat{H}_1^8 \hat{H}_2 \\
& + 36 \hat{H}^2 \hat{H}_1^6 \big(680569 \hat{H} \hat{H}_{4} - 21657984 \hat{H}_2^2\big) \\
& + 18 \hat{H}^3 \hat{H}_1^5 \big(103265 \hat{H} \hat{H}_{5} + 21611188 \hat{H}_{3} \hat{H}_2\big) \\
& + 3 \hat{H}^3 \hat{H}_1^4 \big[191866041 \hat{H}_2^3 
    + \hat{H} \big(11130458 \hat{H}_{3}^2 - 412337 \hat{H} \hat{H}_{6}\big) \\
&\qquad + 13976700 \hat{H} \hat{H}_{4} \hat{H}_2 \big] \\
& + 12 \hat{H}^4 \hat{H}_1^3 \big(8636 \hat{H}_{7} \hat{H}^2 
    - 1749939 \hat{H}_{5} \hat{H} \hat{H}_2 
    - 3497158 \hat{H}_{3} \hat{H}_{4} \hat{H} \\
&\qquad + 22649052 \hat{H}_{3} \hat{H}_2^2 \big) \\
& + \hat{H}^4 \hat{H}_1^2 \Big[133697655 \hat{H}_2^4 
    - 131823180 \hat{H} \hat{H}_{4} \hat{H}_2^2 \\
&\qquad + 12 \hat{H} \big(262657 \hat{H} \hat{H}_{6} - 16541011 \hat{H}_{3}^2) \hat{H}_2 \\
& \qquad + 2 \hat{H}^2 \big(-28914 \hat{H} \hat{H}_{8} 
    + 2806627 \hat{H}_{4}^2 + 4151352 \hat{H}_{3} \hat{H}_{5} \big) \Big] \\
& + \hat{H}^5 \Big[-34502625 \hat{H}_2^5 
    + 9362016 \hat{H} \hat{H}_{4} \hat{H}_2^3 \\
    &\qquad + \hat{H} \big(2316753 \hat{H} \hat{H}_{6} + 76988 \hat{H}_{3}^2 \big) \hat{H}_2^2 \\
& \qquad + \hat{H}^2 \big(-52200 \hat{H} \hat{H}_{8} 
    + 9040251 \hat{H}_{4}^2 + 13154788 \hat{H}_{3} \hat{H}_{5} \big) \hat{H}_2 \\
& \qquad - 2 \hat{H}^2 \Big(53612 \hat{H} \hat{H}_{7} \hat{H}_{3} 
    - 7289089 \hat{H}_{4} \hat{H}_{3}^2 \\
&\qquad\qquad+ \hat{H} \big(623 \hat{H} \hat{H}_{10} + 50882 \hat{H}_{5}^2 + 86869 \hat{H}_{4} \hat{H}_{6} \big) \Big) \Big] \\
& + 2 \hat{H}^5 \hat{H}_1 \Big(8603802 \hat{H} \hat{H}_{5} \hat{H}_2^2 
    - 150844032 \hat{H}_{3} \hat{H}_2^3 \\
&\qquad + 2 \hat{H} \big(48651 \hat{H} \hat{H}_{7} + 17077496 \hat{H}_{3} \hat{H}_{4} \big) \hat{H}_2 \\
& \qquad + \hat{H} \big(-10634 \hat{H}^2 \hat{H}_{9} 
    + 8348120 \hat{H}_{3}^3 
    + 550452 \hat{H} \hat{H}_{6} \hat{H}_{3} \\
&\qquad\qquad+ 1013437 \hat{H} \hat{H}_{4} \hat{H}_{5} \big) \Big)
\Big] \hat{H}^{-3} \d{\eta}.
\end{longalign}